\documentstyle[twoside,fleqn,espcrc2]{article}
\input{psfig}

\newcommand{\AmS}{{\protect\the\textfont2
  A\kern-.1667em\lower.5ex\hbox{M}\kern-.125emS}}

\title{Spin-Parity Analysis of the Centrally produced $K_s K_s$ system at
	800 GeV}

\author{M. A. Reyes\address{Universidad de Guanajuato,
	        Leon, Guanajuato, Mexico},
	M. C. Berisso\address{University of Massachusetts, Amherst,
		Massachusetts, USA},
	D. Christian\address{Fermilab, Batavia, Illinois, USA}, 
	J. Felix$^{\rm a}$,
	A. Gara\address{Columbia University, Nevis Labs, New York, USA},
        E. Gottschalk\address{University of Illinois at Urbana-Champaign,
		Champaign, Illinois, USA},
	\underline{G. Gutierrez}$^{\rm c}$,
	\mbox{E. P. Hartouni$^{\rm b}$},
	B. Knapp$^{\rm d}$,
	M. N. Kreisler$^{\rm b}$,
	S. Lee$^{\rm b}$,
	K. Markianos$^{\rm b}$,
	G. Moreno$^{\rm a}$,
	M. Sosa$^{\rm a}$,
	\mbox{A. Wehmann$^{\rm c}$},
	D. Wesson$^{\rm b}$}
       
\begin{document}

\begin{abstract}
Results are presented of the spin-parity analysis on a sample of centrally
produced mesons in the reaction {$p p \rightarrow p_{slow} (K_s K_s)
p_{fast}$} with 800 GeV protons on liquid hydrogen.  The spin-parity
analysis in the mass region between threshold and 1.58 GeV/$c^2$ shows
that the {$K_s K_s$} system is produced mainly in $S$ wave.  The
{$f_0(1500)$} is clearly observed in this region.  Above 1.58 GeV/$c^2$
two solution are possible, one with mainly $S$ wave and another
with mainly $D$ wave.  This ambiguity prevents a unique determination
of the spin of the {$f_J(1710)$} meson.
\end{abstract}

\maketitle

\section{INTRODUCTION}

The first evidence of the Central Production of $f_0(1500)$ in the reaction
\begin{equation}
p p \rightarrow p_s (K_s K_s) p_f, \hspace{5mm} 
K_s \rightarrow \pi^+ \pi^-
\label{eq:reaction}
\end{equation}
is presented here.  The $f_0(1500)$ was first observed in
$K^- p$~\cite{Asto88} interactions and beautifully confirmed in low energy
$\overline{p} p$ annihilations by the Crystal Barrel
Collaboration~\cite{Amsl95}.  Its properties are of current interest because
it is considered a candidate to be the lowest lying glueball
state~\cite{AmCl96}.

One of the advantages of the final state selected is that only states
with quantum numbers $J^{PC}=(even)^{++}$ are allowed to decay into
$K_s K_s$.  This not only greatly simplifies the analysis but eliminates
confusion coming from all the other states.  The results presented here
are based on 10\% of the 5 x $10^9$ events recorded by FNAL E690 during
Fermilab's 1991 Fixed Target run.

\section{DATA SELECTION}

\subsection{The detector}

The data was taken at Fermilab with an 800 GeV proton beam on a liquid
hydrogen ($LH_2$) target, and the E690 spectrometer. The spectrometer is
composed of two parts: a) the Main Spectrometer (MS), and b) the Beam
Spectrometer (BS).  The MS has an approximately conical geometrical acceptance
with an average 700 mrad radius, good momentum resolution from about 0.2 to 15
GeV/$c$, a Freon 114 threshold Cherenkov counter with a pion threshold of 2.6
GeV/$c$, a time of flight system (TOF) with $\pi/p$ separation up to 1.5
GeV/$c$, and a target veto system.  Neither the TOF nor the Cherenkov counter
were used in the work presented here.

The BS, used to measure the incoming and outgoing protons, has an
approximately conical geometrical acceptance with an average radius of about
1 mrad, a $p_t$ resolution of 6 MeV/$c$, and a longitudinal momentum
resolution of 425 MeV/$c$.  The longitudinal momentum acceptance of the BS for
the interacted beam ranges from approximately 650 to 800 GeV/$c$.

The trigger required an equal number on incoming and outgoing tracks in the BS
and at least one additional track in the MS.

\subsection{Event selection}

\begin{figure}[htb]
%%\framebox[75mm]{\rule{0mm}{70mm}}
 \vspace{-1.5cm}
 \begin{minipage}[t]{7.5cm}
 \hspace*{-.8cm}
 \psfig{figure=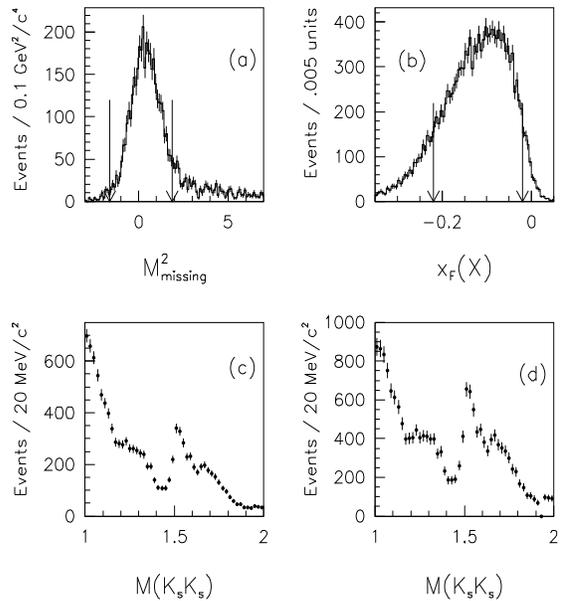,height=19cm,width=17.5cm}
 \end{minipage}
 \vspace*{-10cm}
\caption{a) Missing mass squared for \mbox{$1.4<M(K_s K_s)<1.8$ GeV/$c^2$}.
	 b) Uncorrected $x_F$ distribution.
	 c) Measured $K_s K_s$ invariant mass.
	 d) Acceptance corrected and background subtracted $K_s K_s$
	    invariant mass.}
\label{fig:mass}
\end{figure}

Final state (\ref{eq:reaction}) was selected by requiring a primary vertex in
the $LH_2$ target with two $K_s$, an incoming beam track, and a fast forward
proton.  No direct measurement was made of the slow proton $p_s$. The target
veto system was used to reject events with more than a missing proton.  The
events were accepted when either no veto counter was on, or only one veto
counter was on with the missing $p_t$ pointing to it.  About 12\% of the
selected events were rejected with the veto system. The missing mass squared
seen in Figure~\ref{fig:mass}.a shows a clear proton peak with little
background; the arrows indicate the cuts used in the event selection.

The MS has essentially no acceptance for \mbox{$x_F > 0$}, which insures a gap
of at least 3.5 units of rapidity between $p_f$ and the central products.
The average rapidity gap between the $K_s K_s$ system and $p_s$ is 2.5
units.  Figure~\ref{fig:mass}.b shows the uncorrected $x_F$ distribution
for the $K_s K_s$ system, the arrows indicate the cuts used in the
event selection.

The $\pi^+ \pi^-$ invariant mass for the $K_s$'s has a width of
$\sigma = 2$ MeV/$c^2$.  No direct particle ID (Cherenkov or TOF) was
used to identify the $K_s$ decay products.  In about 7\% of the events a
$K_s$ is compatible with a $\Lambda$; these events were kept, to avoid any
biases in the angular distributions.

The proton mass was assigned to the missing particle in the events that
passed the cuts, then the three momenta of $p_s$ and the longitudinal momentum
of $p_f$ were calculated using energy and momentum conservation.

Figure~\ref{fig:mass}.c shows the $K_s K_s$ invariant mass for the 11182
events that passed the previous cuts and were used in the analysis.  
The analysis was not continued beyond 2 GeV/$c^2$ because the number of events
is very low.  But for $-0.22<x_F<-0.02$ the $K_s K_s$ invariant mass beyond 2
GeV/$c^2$ is smooth, with no evidence of the $\xi(2230)$ state seen by the BES
Collaboration~\cite{BES2pp}.

\section{DATA ANALYSIS}

\subsection{Angular distributions}

\begin{figure}[!bt]
%\framebox[75mm]{\rule{0mm}{155mm}}
 \vspace{-1.5cm}
 \begin{minipage}[t]{7.5cm}
 \hspace*{-1cm}
 \psfig{figure=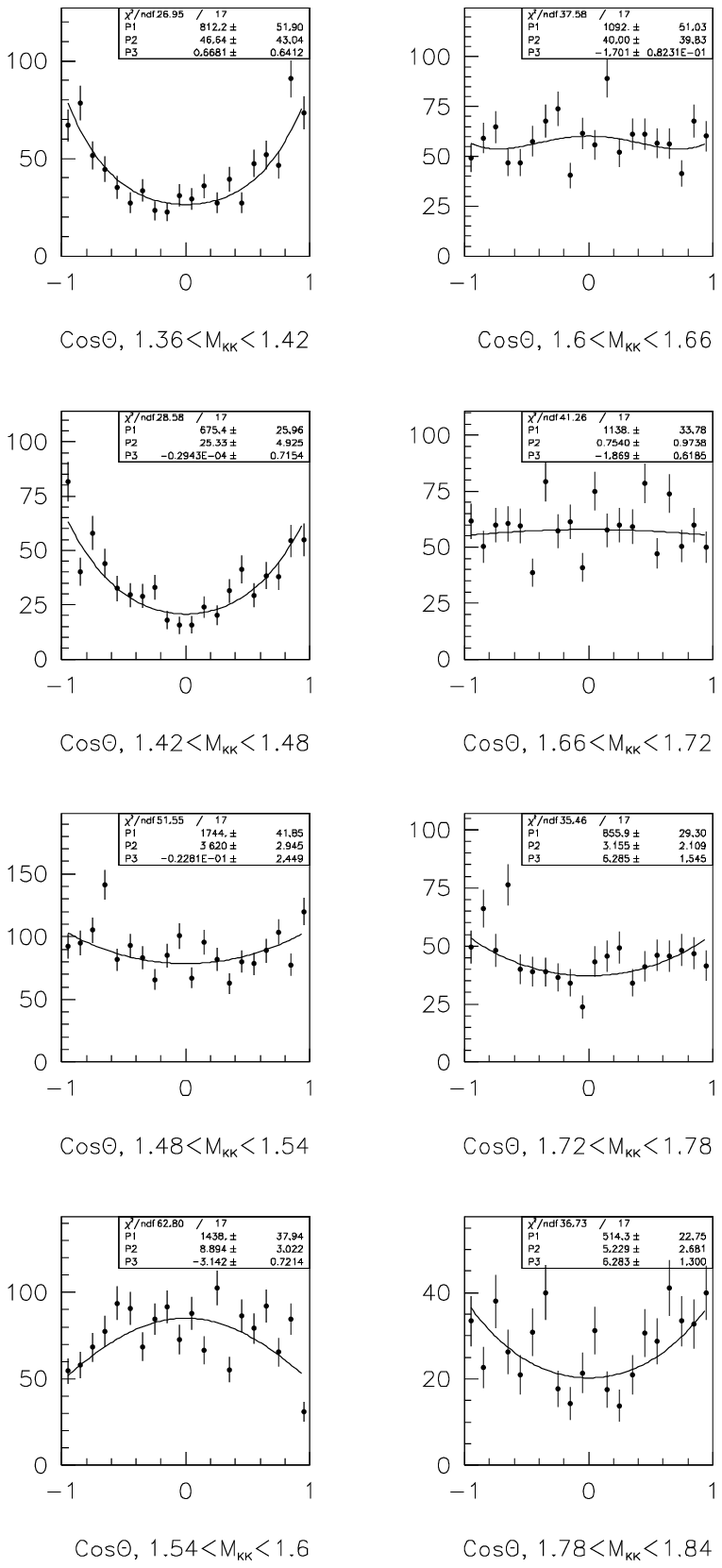,height=18.5cm,width=17.5cm}
 \end{minipage}
 \vspace*{-1.8cm}
\caption{Acceptance corrected $\cos\theta$ angular distributions in bins
	of the $K_s K_s$ invariant mass, starting at 1.36 GeV/$c^2$ in
	steps of 60 MeV/$c^2$.}
\label{fig:cost}
\end{figure}

\begin{figure}[!bt]
%\framebox[75mm]{\rule{0mm}{155mm}}
 \vspace{-1.5cm}
 \begin{minipage}[t]{7.5cm}
 \hspace*{-1cm}
 \psfig{figure=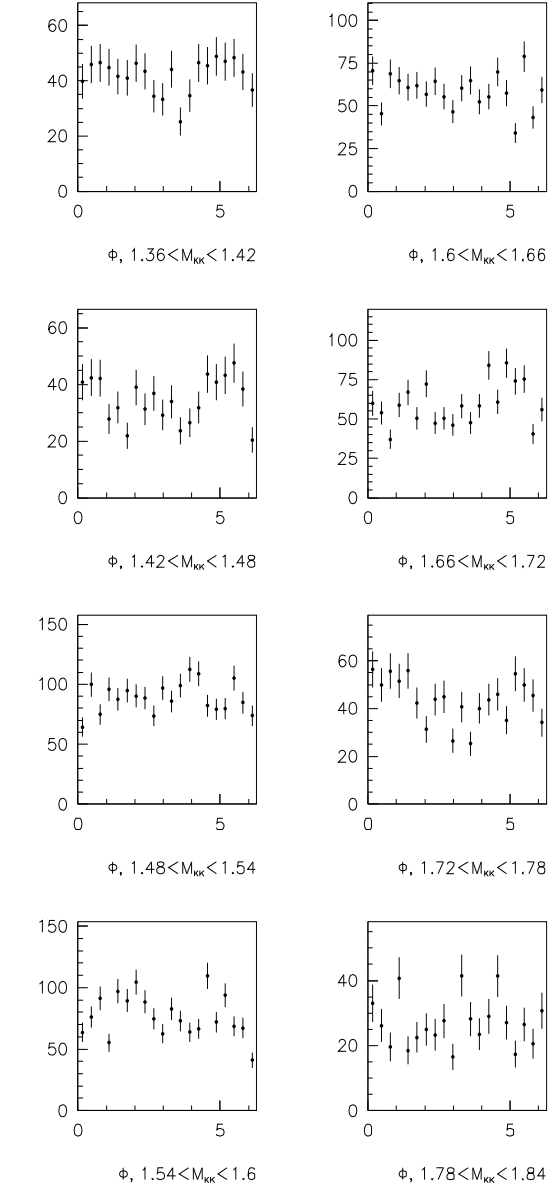,height=18.5cm,width=17.5cm}
 \end{minipage}
 \vspace*{-1.8cm}
\caption{Acceptance corrected $\phi$ angular distributions in bins of the
	$K_s K_s$ invariant mass, starting at 1.36 GeV/$c^2$ in steps of
	60 MeV/$c^2$.}
\label{fig:phi}
\end{figure}

The reaction studied here was analyzed as a two step process: the
production step in which an $(X)$ system is formed by the collision of
two objects (from now on referred to as ``pomerons") emitted by each of the
scattered protons, and the decay step in which the object $(X)$ decays into
$K_s K_s$. The production coordinate system was defined in the
CM of the $(X)$ system, with the y-axis perpendicular to the plane of the two
``pomerons" in the overall CM, and the z-axis in the direction of the ``beam
pomeron" in the $(X)$ CM. The two variables needed to specify the decay
process were taken as the polar and azimuthal angles $(\theta,\phi)$ of one of
the $K_s$ (taken at random) in the production coordinate system.  The
acceptance corrected $\cos\theta$  and $\phi$ distributions are shown in
Figures~\ref{fig:cost} and ~\ref{fig:phi}. The acceptance is flat in $\phi$,
and dips near $\cos\theta=\pm 1$. On average the correction at $\cos\theta=\pm
1$, relative to the correction at 0, is 65\%.

The five variables used to specify the production process were the transverse
momenta of the slow and fast protons ($p_{t,s}^2, p_{t,f}^2$), the $x_F$ and
invariant mass of the $K_s K_s$ system, and $\delta$, the angle between the
planes of the scattered protons in the $K_s K_s$ CM.  Although our 11182
events constitute a large sample, it is not large enough to bin the data in
all five production variables.   The present analysis was done in bins of the
$K_s K_s$ invariant mass for the $x_F$ selected region, and integrating over
$p_{t,s}^2$, $p_{t,f}^2$ and $\delta$.

\subsection{Partial wave analysis}

\begin{figure}[!bt]
%\framebox[75mm]{\rule{0mm}{148mm}}
 \vspace{-1.5cm}
 \begin{minipage}[t]{7.5cm}
 \hspace*{-.8cm}
 \psfig{figure=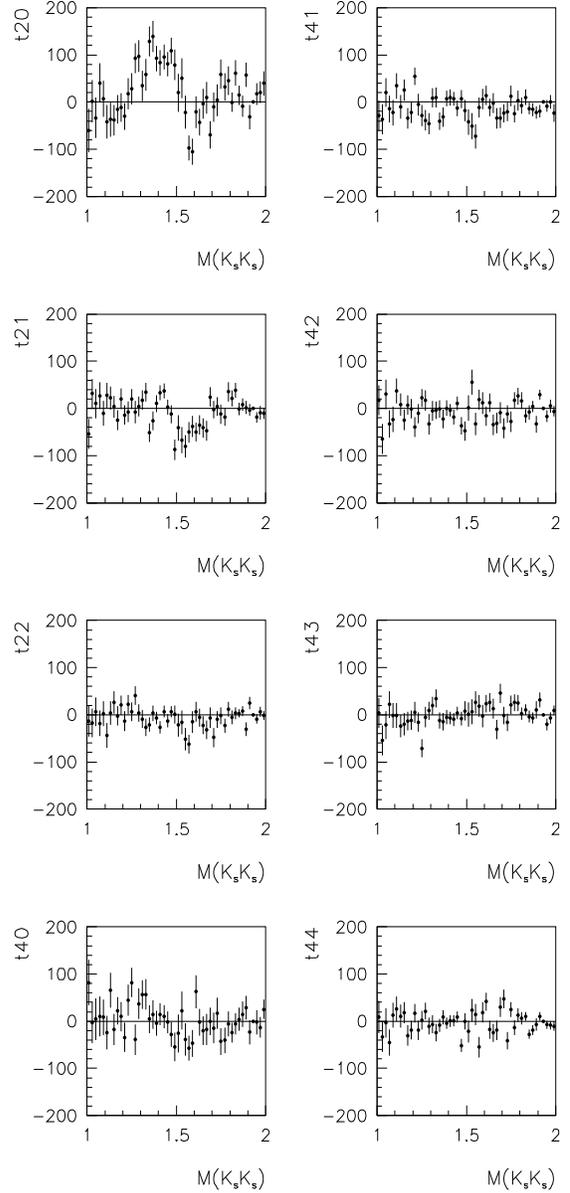,height=18.5cm,width=17.5cm}
 \end{minipage}
 \vspace*{-1.8cm}
\caption{Acceptance corrected moments as a function of the $K_s K_s$
	 invariant mass.}
\label{fig:moments}
\end{figure}

The acceptance corrected moments, defined as
\begin{equation}
I(\Omega)=\frac{1}{\sqrt{4\pi}}\{\sum_{l}t_{l0}Y_l^0+
          2\sum_{l,m>0}t_{lm}Re(Y_l^m)\}
\end{equation}
are shown in Figure~\ref{fig:moments}.  The odd moments (not shown) are
consistent with zero, as expected for a system of two identical bosons.  The
acceptance corrected mass distribution ($t_{00}$ moment) is shown in 
Figure~\ref{fig:mass}.d.  The error bars are statistical errors only.

In the two step process considered here the $(X)$ system is formed
by the interchange of two ``pomerons" and it decays afterwards independently
of the two final state protons.  The two ``pomerons" form a plane; parity
in the strong interactions implies that reflection in this plane should
be a symmetry of the system~\cite{SUrefl}.  Therefore the amplitudes used for
the Partial Wave Analysis (PWA) were defined in the reflectivity
basis~\cite{SU&Tru}.  Since the $t_{43}$ and $t_{44}$ moments are consistent
with zero (see Fig~\ref{fig:moments}), only spherical harmonics with $l=0,2$
and $m=0,\pm1$ were considered.  The waves used were $L_m^\epsilon$, with 
$L=S,D$, $m\ge0$ and $\epsilon=\pm1$:

\begin{equation}
S_0^-=Y_0^0=\frac{1}{\sqrt{4\pi}}
\label{eq:s0}
\end{equation}

\begin{equation}
D_0^-=Y_2^0=\sqrt{\frac{5}{16\pi}}~(3\;\cos^2\theta - 1)
\label{eq:d0}
\end{equation}

\begin{equation}
D_1^-=\frac{Y_2^1-Y_2^{-1}}{\sqrt{2}}=-\sqrt{\frac{15}{16\pi}}~\sin2\theta
	~\cos\phi
\label{eq:d-}
\end{equation}

\begin{equation}
D_1^+=\frac{Y_2^1+Y_2^{-1}}{\sqrt{2}}=-i\sqrt{\frac{15}{16\pi}}~\sin2\theta
	~\sin\phi
\label{eq:d+}
\end{equation}
Waves with different reflectivity $\epsilon$ do not interfere.
\begin{figure}[!bt]
%\framebox[75mm]{\rule{0mm}{148mm}}
 \vspace{-1.5cm}
 \begin{minipage}[t]{7.5cm}
 \hspace*{-.8cm}
 \psfig{figure=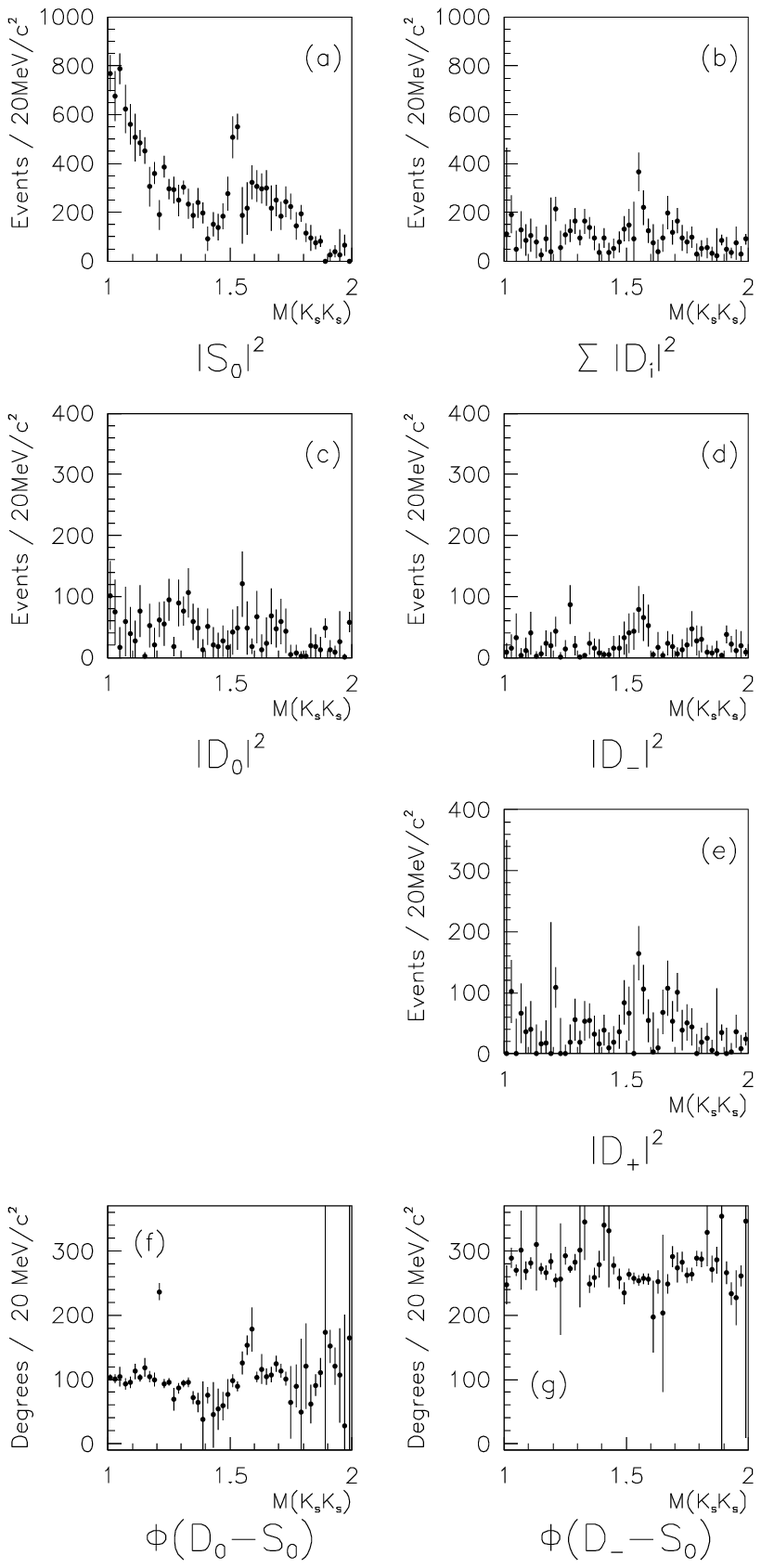,height=18.5cm,width=17.5cm}
 \end{minipage}
 \vspace*{-1.8cm}
\caption{Waves as a function of $K_s K_s$ invariant mass for solution one.
	a) $S$ and b) total $D$ waves, c) to e) individual $D$ wave, and
	f) and g) phases relative to the $S$ wave.}
\label{fig:sol1}
\end{figure}
\begin{figure}[!bt]
%\framebox[75mm]{\rule{0mm}{148mm}}
 \vspace{-1.5cm}
 \begin{minipage}[t]{7.5cm}
 \hspace*{-.8cm}
 \psfig{figure=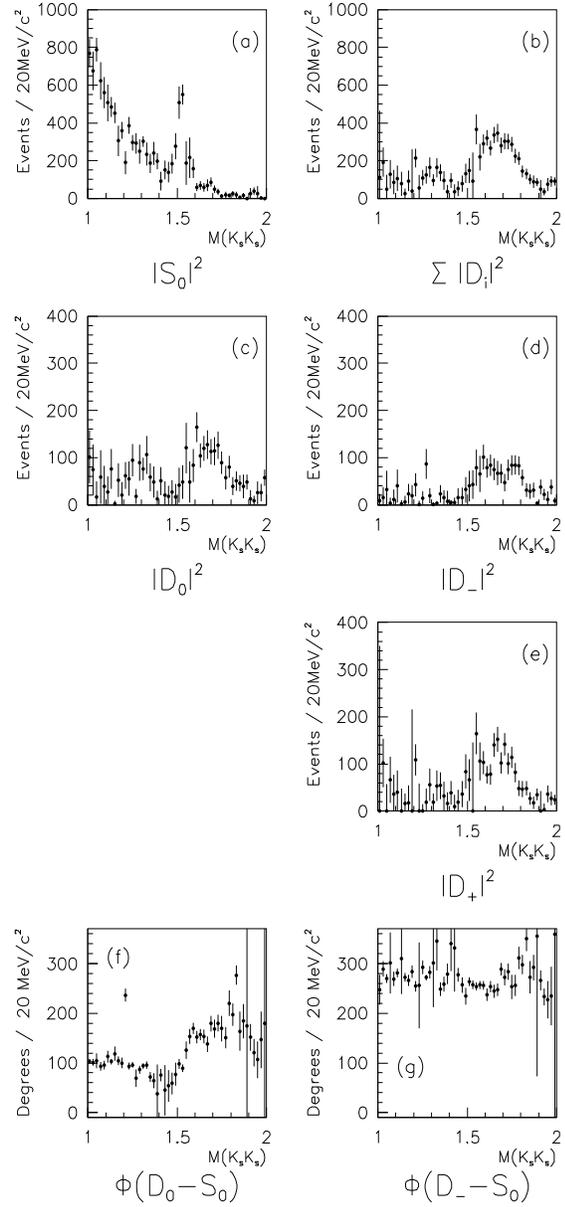,height=18.5cm,width=17.5cm}
 \end{minipage}
 \vspace*{-1.8cm}
\caption{Waves as a function of $K_s K_s$ invariant mass for solution two.
	a) $S$ and b) total $D$ waves, c) to e) individual $D$ wave, and
	f) and g) phases relative to the $S$ wave.}
\label{fig:sol2}
\end{figure}

The PWA analysis was done in two different ways.  First since the $\phi$
angular distributions are fairly flat only $S_0^-$ and $D_0^-$ waves were
used: a) by fitting to the $\cos\theta$ angular distributions, and b) 
by using the extended maximum likelihood method.  The results of the fit to
the $\cos\theta$ angular distributions are shown in Figure~\ref{fig:cost}.
Within errors the results were the same in both cases, giving a solution that,
except for two small $D$ wave contributions at $\sim$1.3 GeV/$c^2$ and 
$\sim$1.6 GeV/$c^2$, was all $S$ wave.  Second, all four waves
(\ref{eq:s0}-\ref{eq:d+}) were used.  The amplitudes were extracted both a)
from the moments shown in Figure~\ref{fig:moments}, and b) by maximizing the
extended likelihood with respect to the four wave moduli and the two relative
phases $\varphi(D_{0,1}^-)-\varphi(S_0^-)$.  Within errors both analyses gave
the same answer.

When using the four waves (\ref{eq:s0}-\ref{eq:d+}) the inherent ambiguities
of a two body system are such that there are two solutions for each mass
bin~\cite{SU&Tru,Barrel}. Both solutions give identical moments or identical
values of the Likelihood. In order to continue the solutions from one mass bin
to the next, one follows the Barrelet zeros.  In general these zeros are
complex and one lies above the real axis and the other lies below it.  When
the zeros cross the real axis the solutions bifurcate~\cite{SU&Tru,Barrel}. 
In the analysis presented here, there is a bifurcation point at 1.58
GeV/$c^2$.  Before this bifurcation point there are only two solutions, one
which is mostly $S$ wave, and another that is mostly $D$ wave.  Since at
threshold the $K_s K_s$ cross section is dominated by the presence of the
$f_0(980)$~\cite{Morgan} it is possible to eliminate the solution that has a
very small $S$ wave contribution at threshold.  The remaining solution
bifurcates at 1.58 GeV/$c^2$ into a solution that has a large $S$ wave
contribution (solution one), and another that has a large $D$ wave component
(solution two).  The solutions obtained using maximum likelihood are shown in
Figures~\ref{fig:sol1} and~\ref{fig:sol2}.  Solution one is shown in
Figure~\ref{fig:sol1}, and solution two in Figure~\ref{fig:sol2}. The errors
shown are statistical errors only.

A striking feature of both solutions is the large $S$ wave peak observed at
$\sim$1.5 GeV/$c^2$. This corresponds to the $f_0(1500)$ observed by the
Crystal Barrel collaboration~\cite{Amsl95}. The mass peaks at 1.52 GeV/$c^2$
instead of at 1.50 GeV/$c^2$, but this could be easily due to interference
with the $S$ wave background.

Beyond 1.58 GeV/$c^2$ both solution one and two are equally valid, and at
the moment there is no way to decide with this data alone which of the two
solutions is the correct one.  However, given that beyond 1.58 GeV/$c^2$ the
angular distributions are fairly structureless, and that an analysis in
$\cos\theta$ alone gives very little $D$ wave, solution one could be favored.

\section{CONCLUSIONS}

A PWA analysis in a sample of 11182 centrally produced $K_s K_s$ events
at 800 GeV has been presented.  Two solutions have been found in the analysis.
In both of them a clear $f_0(1500)$ has been observed.  The ambiguity above
1.58 GeV/$c^2$ prevents a unique determination of the spin of the $f_J(1710)$
meson.  Due to lack of statistics the analysis was not carried out beyond 
2 GeV/$c^2$, but the $K_s K_s$ invariant mass spectrum is smooth beyond
that point and shows no sign of the $\xi(2230)$ meson seen by the BES
Collaboration~\cite{BES2pp}.

\section{ACKNOWLEDGMENTS}

G. Gutierrez wishes to thank many people for useful discussions during the
Conference, the organizers for the invitation, and S.U. Chung for useful
discussions and advice during this work.  M. A. Reyes wishes to thank the
support provided by Academia de la Investigacion Cientifica (Mexico), and the
USA-Mexico Foundation for Science.

\end{document}